\documentclass[prl,twocolumn,superscriptaddress,showpacs]{revtex4}
\usepackage{amsmath}
\usepackage{amsfonts}
\usepackage{amssymb}
\usepackage{graphicx}

\begin{document}

\title{Macroscopic Thermodynamical Witnesses of Quantum Entanglement}
\author{{\v C}aslav Brukner}
\email{caslav.brukner@univie.ac.at} \affiliation{Optics Section,
The Blackett Laboratory, Imperial College, Prince Consort Road,
London, SW7 2BW, United Kingdom} \affiliation{Institut f\"ur
Experimentalphysik, Universit\"at Wien, Boltzmanngasse 5, A--1090
Wien, Austria}
\author{Vlatko Vedral} \email{v.vedral@imperial.ac.uk} \affiliation{Optics Section,
The Blackett Laboratory, Imperial College, Prince Consort Road,
London, SW7 2BW, United Kingdom}

\date{\today}

\begin{abstract}

We show that macroscopic thermodynamical properties - such as
functions of internal energy and magnetization - can detect
quantum entanglement in solids at nonzero temperatures in the
thermodynamical limit. We identify the parameter regions (critical
values of magnetic field and temperature) within which
entanglement is witnessed by these thermodynamical quantities.

\end{abstract}

 \maketitle

An entangled quantum system is impossible to describe by the
states of its (local) constituents alone~\cite{schroedinger}. This
fundamental feature of quantum mechanics can be manifested in a
phenomenon known as quantum non-locality~\cite{bell}. Besides
being of fundamental interest, entanglement is considered as the
crucial resource for quantum information
processing~\cite{nielsenchuang}, but its effects are not generally
seen beyond the atomic scale and only in well controlled
laboratory conditions. It is an intriguing and fascinating
question to what extend entanglement may develop naturally in
realistic complex systems and can affect their macroscopic
properties.

Recently, extensive efforts have been made to understand
theoretically and to quantify entanglement in solid state
systems~\cite{nielsen,arnesen,gunlycke,oconnor,wang1,wang2,osborne,osterloh,vidal,lattore,jin,asoudeh,vlatko,vlatkosuperconducter,fan,ghosh}.
Within various models of interacting spins in arrays entanglement
was found to be present at moderate nonzero temperatures --
phenomenon known as \textit{thermal
entanglement}~\cite{nielsen,arnesen} -- and has been linked to the
existence of critical
phenomena~\cite{osborne,osterloh,vidal,lattore,sachdev}. Most of
the studies, however, were limited to only small number of spin
sites. On the other hand, in the thermodynamical limit (where the
number $N$ of spins tends to infinity) entanglement was
predominately investigated for the systems in their ground
states~\cite{oconnor,osborne,osterloh,vidal,lattore,jin}.

Only very few results are known about a possibility of existence
of entanglement and its properties in the thermal states in the
thermodynamical limit. Entanglement as measured by
concurrence~\cite{wootters} was shown to exist in this limit at
nonzero temperatures in the transverse Ising model~\cite{osborne}
and the antiferromagnetic Heisenberg model~\cite{wang1,wang2} and
an upper bound for the multipartitive entanglement in the
transverse Ising model was given~\cite{vlatko}. Finally, it was
suggested that macroscopic entanglement is possible even at high
temperatures (as high as 160 Kelvin) in high-temperature
superconducters~\cite{vlatkosuperconducter} (see also
Ref.~\cite{fan}).

Thermodynamic laws are of very general validity, and they do not
depend on the details of the interactions or type of the
(microscopic) constituents of the system being studied. When the
system is at thermal equilibrium under a certain temperature $T$,
it is in a thermal state $\rho\!=\!e^{-H/kT}/Z$, where
$Z\!=\!\text{Tr}(e^{-H/kT})$ is the partition function, $H$ is the
Hamiltonian and $k$ is the Boltzmann constant. The partition
function is the central object of statistical physics from which
all other thermodynamical quantities can be derived: e.g. internal
energy $U\!=\!-(1/Z)(\partial Z/\partial\beta)$ or magnetization
$M\!=\!-(1/Z\beta)(\partial Z/\partial B)$, where
$\beta\!=\!1/kT$. It is generally believed that although the
partition function is determined by the eigenvalues of Hamiltonian
only, detecting entanglement in solids requires in addition the
knowledge of the energy eigenstates. This is in general a hard
problem and origin of the main difficulties in the research on
entanglement in solid state systems.

Recently, however, it was demonstrated experimentally that
entanglement can affect macroscopic properties of solids, albeit
at very low (critical) temperature (below 1 Kelvin)~\cite{ghosh}.
This extraordinary result opens up a possibility that purely
quantum correlations between microscopic constituents of the solid
may be detected by only a small number of macroscopic
thermodynamical properties. In the similar spirit concurrence in
the isotropic XXX Heisenberg model~\cite{wang1,wang2} and an upper
bound on entanglement in the transverse Ising model~\cite{vlatko}
was given in terms of internal energy.

In this Letter we show that thermodynamical macroscopic properties
can serve as (multipartitive) entanglement witnesses for the
thermal states in the thermodynamical limit. Entanglement
witnesses are observables which have positive expectation values
for separable states and negative one for some, specific,
entangled states~\cite{horodecki}. We show that a function of
internal energy and magnetization can detect entanglement in the
thermal states in the (XX and XXX) Heisenberg models for a wide
range of values of the external magnetic field and temperature.
The critical temperature below which entanglement is present in
the system is of the order of the coupling constant $J$
(Experimental estimations of this constant gives the values as
high as about 10 Kelvin~\cite{experiment}, where $J$ is measured
in $kT$ units. Compare with Ref.~\cite{vlatkosuperconducter}).

We consider the linear (1D) spin chain in the Heisenberg model.
The interaction is between nearest-neighbor spins and the external
field $B$ is along $z$ direction. The Hamiltonian is given by
\begin{equation}
H=-\sum_{i=1}^{N} (J_x \sigma_{i}^{x} \sigma_{i+1}^{x} + J_y
\sigma_{i}^{y} \sigma_{i+1}^{y} + J_z \sigma_{i}^{z}
\sigma_{i+1}^{z}) - B \sum_{i=1}^{N} \sigma^z_i,
\label{hamiltonian}
\end{equation}
where $\sigma^x_i$, $\sigma^y_i$, and $\sigma ^z_i$ are the Pauli
spin operators for the $i$th spin. Throughout the paper we will
consider special cases of this Hamiltonian -- the isotropic XXX
Heisenberg model with $J_x\!=\!J_y\!=\!J_z\!=\!J$ and isotropic XX
Heisenberg model with $J_x\!=\!J_y\!=\!J$ and $J_z\!=\!0$. The
regimes $J>0$ and $J<0$ correspond to the antiferromagnetic and
the ferromagnetic cases, respectively.

We will now derive a thermodynamical entanglement witness for a
solid state system in thermal equilibrium. Using $U\!=\!\langle H
\rangle$ and $M\!=\! \sum_{j=1}^{N} \langle \sigma^z_j \rangle$ we
obtain
\begin{equation} \frac{U+BM}{NJ}=-\frac{1}{N} \sum_{i=1}^{N}
(\langle \sigma_{i}^{x} \sigma_{i+1}^{x} \rangle + \langle
\sigma_{i}^{y} \sigma_{i+1}^{y} \rangle + \langle \sigma_{i}^{z}
\sigma_{i+1}^{z}\rangle) \label{okako}
\end{equation}
from Eq. (\ref{hamiltonian}). The right-hand of Eq. (\ref{okako})
is an entanglement witness as shown in Ref.~\cite{geza}: for any
separable state, that is, for any classical mixture of the
products states: $\rho=\sum_k w_k \rho^1_k \otimes \rho^2_k
\otimes ... \otimes \rho^N_k$, one has
\begin{equation}
\frac{1}{N} |\sum_{i=1}^{N} (\langle \sigma_{i}^{x}
\sigma_{i+1}^{x} \rangle + \langle \sigma_{i}^{y} \sigma_{i+1}^{y}
\rangle + \langle \sigma_{i}^{z} \sigma_{i+1}^{z}\rangle)| \leq 1.
\label{vudubluz}
\end{equation}
The proof is based on the fact that for any product state
$\rho^1_k \otimes ... \otimes \rho^N_k$ and for every $i$ one has
$ |\langle \sigma_{i}^{x} \sigma_{i+1}^{x} \rangle + \langle
\sigma_{i}^{y} \sigma_{i+1}^{y} \rangle + \langle \sigma_{i}^{z}
\sigma_{i+1}^{z}\rangle| = |\langle \sigma_{i}^{x} \rangle \langle
\sigma_{i+1}^{x} \rangle + \langle \sigma_{i}^{y} \rangle \langle
\sigma_{i+1}^{y} \rangle + \langle \sigma_{i}^{z} \rangle \langle
\sigma_{i+1}^{z}\rangle| \leq \sqrt{\langle \sigma_{i}^{x}
\rangle^2 + \langle \sigma_{i}^{y} \rangle^2 + \langle
\sigma_{i}^{z} \rangle^2} $ $ \sqrt{\langle \sigma_{i+1}^{x}
\rangle^2 + \langle \sigma_{i+1}^{y} \rangle^2 + \langle
\sigma_{i+1}^{z} \rangle^2} \leq 1$. This is also valid for any
convex sum of product states (separable states). The upper bound
was found by using the Cauchy-Schwarz inequality and knowing that
for any state $\langle \sigma^{x} \rangle^2 + \langle \sigma^{y}
\rangle^2 + \langle \sigma^{z} \rangle^2\leq 1$. It is important
to note that the same proof can also be applied if one considers
XX Heisenberg model. In this case one has
\begin{equation}
\frac{1}{N} |\sum_{i=1}^{N} (\langle \sigma_{i}^{x}
\sigma_{i+1}^{x} \rangle + \langle \sigma_{i}^{y} \sigma_{i+1}^{y}
\rangle| \leq 1 \label{igrajvudu}
\end{equation}
for any separable state.

We now give our thermodynamical entanglement witness: if, in the
isotropic XXX or XX Heisenberg model, one has
\begin{equation}
\frac{|U+BM|}{N|J|} > 1, \label{niko}
\end{equation}
then the solid state system is in an entangled state. The
entanglement witness is physically equivalent to the exchange
interaction energy or, equivalently, to the difference between the
total (internal) energy $U$ and the magnetic energy $-BM$. From
the tracelessness of the Pauli operators one can easily see that
$\lim_{T\rightarrow \infty} U \rightarrow 0$. This means that the
value of the internal energy as given by (\ref{niko}) should be
defined relatively to the referent value of zero energy in the
limit of high temperatures.

We now give an explicit example of a state that violates
Ineqs.~(\ref{vudubluz}) and (\ref{igrajvudu}). This then completes
the proof that expression (\ref{niko}) is indeed an entanglement
witness and not just a bound that is trivially satisfied by any
quantum state. As an example of such a state we take the ground
state of the antiferromagnetic isotropic XXX Heisenberg model with
zero magnetic field. The energy of this state was found to
be~\cite{hulthen,oconnor}: $|E_0/JN|\!=\!1/N|\sum_{i=1}^N (\langle
\sigma_{i}^{x} \sigma_{i+1}^{x} \rangle_0 + \langle \sigma_{i}^{y}
\sigma_{i+1}^{y} \rangle_0 + \langle \sigma_{i}^{z}
\sigma_{i+1}^{z}\rangle_0)|\!=\!1.773 \!>\! 1$, where the index
``0" denotes that the mean values are taken for the ground state.
Furthermore, due to the symmetry of the XXX Heisenberg Hamiltonian
one has $ E_0/(3NJ)\!=\!\langle \sigma_{i}^{x} \sigma_{i+1}^{x}
\rangle_0 \!=\! \langle \sigma_{i}^{y} \sigma_{i+1}^{y} \rangle_0
\!=\! \langle \sigma_{i}^{z} \sigma_{i+1}^{z}\rangle_0 \!=\!
-1.773/3$ for every $i$. This implies that $1/N|\sum_{i=1}^N
(\langle \sigma_{i}^{x} \sigma_{i+1}^{x} \rangle_0 + \langle
\sigma_{i}^{y} \sigma_{i+1}^{y} \rangle_0)|=$ $ 1.182 > $ $ 1$.
Therefore, Ineq.~(\ref{niko}) is an entanglement witness for the
solid state systems described by XXX or XX Heisenberg interaction.

We will now discuss various concrete models of spin interaction of
which some are exactly solvable and for which dependence of
internal energy $U$ and magnetization $M$ on temperature $T$ and
magnetic field $B$ are known. This will help us to determine the
parameter regions of $T$ and $B$ within which one has entanglement
in the solids.

We first consider \textit{XXX Heisenberg model with no magnetic
field} $(J_x\!=\!J_y\!=J_z\!=\!J$ and $B\!=\!0)$. Since there is
no magnetic field, symmetry requires that magnetization vanishes
and thermodynamical witness (\ref{niko}) reduces to
\begin{equation}
\frac{|U|}{N|J|} > 1. \label{kao}
\end{equation}
In Ref.~\cite{wang1} it was shown that concurrence $C$ (as a
measure of bipartite entangelment~\cite{wootters}) is zero at any
temperature in the ferromagnetic case and that it is given by
$C=\frac{1}{2} \text{ max }\left[0,|U|/(NJ)-1\right]$ in the
antiferromagnetic case. Thus $C$ is nonzero if and only if
$|U|/(NJ)>1$. This shows that our thermodynamical entanglement
witness can detect entire bipartite entanglement as measured by
concurrence. Furthermore, the fact that the value of the
entanglement witness for the ground state is well above the limit
of $1$ $(|E_0/JN|=1.773)$ suggests that entanglement could exist
and be detected by the thermodynamical witness at nonzero
temperatures as well.

Next we analyze \textit{XXX Heisenberg model with nonzero magnetic
field} $(J_x\!=\!J_y\!=J_z\!=\!J$ and $B\!\neq\!0)$. Recently, it
was shown that concurrence vanishes in the thermodynamical limit
for the ferromagnetic case $(J<0)$ at low temperatures when only
ground state and the first excited states (single spin
excitations) are populated~\cite{asoudeh}. Within the validity of
this approximation the partition function is given
by~\cite{asoudeh} $Z\!=\!e^{B\beta(J+B)}(1+e^{-2\beta
B}N/\sqrt{8\pi\beta J})$. Using this we obtain $|U+BM|/(NJ)\!=\!1$
and thus no entanglement can be detected at the level  of
approximation in agreement with the result of Ref.~\cite{asoudeh}.

We proceed with the consideration of \textit{XX Heisenberg model
with nonzero magnetic field} $(J_x\!=\!J_y\!=\!J$, $J_z\!=\!0$ and
$B\! \neq \!0)$. This case is the most interesting as it is
exactly solvable and the partition function was found in
Ref.~\cite{katsura}. Let us introduce the following dimensionless
quantities: $C=B/kT$ and $K=J/kT$ (note a difference of factor 2
in the definitions of $J$ and $K$ with respect to
Ref.~\cite{katsura}) and the function
\begin{equation}
f(K,C,\omega)= \sqrt{2K^2+2K^2\cos{2\omega}-4CK\cos{\omega}+C^2}
\end{equation}
for convenience. Then the internal energy is given
by~\cite{katsura}
\begin{equation}
\frac{U}{N} = -\frac{kT}{\pi} \int_{0}^{\pi} f(K,C,\omega)
\tanh{f(K,C,\omega)} d\omega, \label{energy}
\end{equation}
and the magnetization by~\cite{katsura}
\begin{equation}
\frac{M}{N} = -\frac{1}{\pi} \int_{0}^{\pi}
\frac{4K^2\cos^2{\omega}}{f(K,C,\omega)} \tanh{f(K,C,\omega)}
d\omega \label{magnetization}
\end{equation}
both in ferromagnetic and antiferromagnetic case.

\begin{figure}
\centering
\includegraphics[angle=0,width=6.1cm]{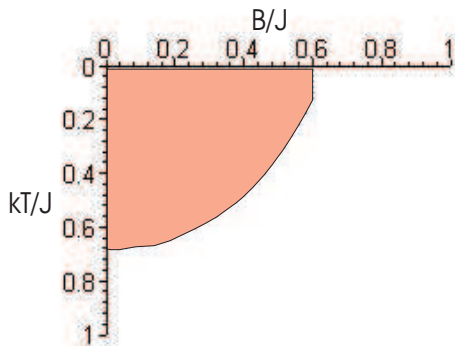}
\caption{The parameter regions of temperature $kT/|J|$ and
external magnetic field $B/|J|$ (expressed in the units of the
coupling constant $J$; $k$ is the Boltzmann constant) where
thermodynamical entanglement witness $|(U+MB)/(NJ)|>1 $ detects
entanglement in the XX Heisenberg solid state system both in
ferromagnetic and antiferromagnetic case.} \label{entwitness}
\end{figure}

We use Eqs.~(\ref{energy}) and (\ref{magnetization}) to determine
the parameter regions of temperature $kT$ and magnetic field  $B$
for which entanglement exists in the solid state system
(Fig.~\ref{entwitness}). The critical values of $kT$ and $B$ below
which entanglement can be detected is of the order of $J$, which
can be as high as $10$ Kelvin~\cite{experiment}.

Finally, we consider \textit{XX Heisenberg model with no magnetic
field} $(J_x\!=\!J_y\!=\!J$, $J_z\!=\!0$ and $B\!=\!0$). This
model is important for quantum information processing as it
describes the effective interaction in cavity QED and between two
quantum dots~\cite{imamoglu}. It also can be used to construct the
controlled-NOT gate~\cite{imamoglu}. Since there is no magnetic
field, magnetization vanishes and thermodynamical witness reduces
again to Eq.~(\ref{kao}). Nevertheless, the expressions
(\ref{energy}) for internal energy and (\ref{magnetization}) for
magnetization are valid also for the case with no magnetic field.
Thus the axes $B=0$ on Fig.~\ref{entwitness} gives the temperature
interval within which entanglement is present in the XX system.

The common feature in all cases for which the existence of
entanglement could here be proven is that both high temperatures
and high values of magnetic field move the thermal states away
from the region with non-zero entanglement. This is understandable
because high values of magnetic field tend to order all spins
parallel to the field which corresponds to an overall state being
a product of the individual spin states. Increasing the
temperature has also entanglement destructive character due to
thermal fluctuations (which, on the other hand, can be understood
due to the interaction between the environment and the solid state
system).

We note that our method for determining entanglement in solids
within the models of Heisenberg interaction is useful in the cases
where other methods fail due to incomplete knowledge of the
system. This is the case when only the eigenvalues but not
eigenstates of the Hamiltonian are known (which is the most usual
case in solid state physics) and thus no measure of entanglement
can be computed. Furthermore, in the cases where we lack complete
description of the systems we can approach the problem
experimentally and determine the value of the thermodynamical
entanglement witness by performing appropriate measurements.

Our work raises a number of interesting questions and
possibilities for generalizations such as consideration of
Hamilonians with higher spins, two- and three-dimensional systems,
non-nearest interactions, anisotropies, other thermodynamical
properties (e.g. heat capacity, magnetic susceptibility) and so
on.

In conclusion, we show that the presence of entanglement in solid
state systems at nonzero temperatures in the thermodynamical limit
can be detected by measuring solely macroscopic thermodynamical
properties. The parameter regions of temperature and magnetic
field are determined for which there is entanglement in the
systems. Besides being of fundamental interest, we expect our
results to be useful for potential physical realization of a
future quantum computer. In the opinion of many researchers, if
the future computer is supposed to reach the stage of wide
commercial application, it should be based on solid states
systems. It will thus be important to derive the critical values
of physical parameters (e.g. high-temperature limit) above which
one can not harness quantum entanglement in solids as a resource
for quantum information processing.

\begin{acknowledgments}

{\v C}.B. has been supported by the European Commission, Marie
Curie Fellowship, Project No. 500764 and by the Austrian Science
Foundation (FWF) Project No. F1506. V.V. acknowledges EPSRC for
financial support.

\end{acknowledgments}

\end{document}